\documentclass[aps,pra,reprint,onecolumn,notitlepage,eqsecnum,floatfix,showkeys,nofootinbib]{revtex4-2}
\usepackage{amsmath}
\usepackage{amsbsy}
\usepackage{amsfonts}
\usepackage{bigints}
\usepackage{xcolor}
\usepackage{graphicx}
\usepackage{hyperref}
\usepackage{multirow}
\usepackage{linearb}
\usepackage{relsize}
\usepackage{amsthm}
\usepackage{fancyvrb}
\usepackage{calrsfs}
\DeclareMathAlphabet\mathbfcal{OMS}{cmsy}{b}{n}

\newcommand{\bq}{\begin{eqnarray}}
\newcommand{\eq}{\end{eqnarray}}
\newcommand{\bqn}{\begin{eqnarray*}}
\newcommand{\eqn}{\end{eqnarray*}}
\newcommand{\bqs}{\begin{subequations}}
\newcommand{\eqs}{\end{subequations}}
\newcommand{\bw}{\begin{widetext}}
\newcommand{\ew}{\end{widetext}}

\newcommand{\rr}{{\boldsymbol r}}

\newcommand{\uu}{{\boldsymbol u}}

\newcommand{\call}{{\cal L}}

\newcommand{\calo}{{\cal O}}

\newcommand{\calt}{{\cal T}}

\newcommand{\calv}{{\cal V}}
\newcommand{\calw}{{\cal W}}
\newcommand{\calk}{{\cal K}}

\SaveVerb{floor}=FLOOR=
\SaveVerb{int}=INT=
\SaveVerb{nint}=NINT=
\SaveVerb{sgn}=SGN=
\SaveVerb{abs}=ABS=

\begin{document}
%%%%%%%%%%%%%%%%%%%%%%%%%%%%%%%%%%%%%%%%%%%%%%%%%%%%%%%%%%%%%%%%%%%%%%%%%%%%%%
%%%%%%%%%%%%%%%%%%%%%%%%%%%%%%%%%%%%%%%%%%%%%%%%%%%%%%%%%%%%%%%%%%%%%%%%%%%%%%
%%%%%%%%%%%%%%%%%%%%%%%%%%%%%%%%%%%%%%%%%%%%%%%%%%%%%%%%%%%%%%%%%%%%%%%%%%%%%%
\title{Electron-Ion Path Integral Monte Carlo with Hard Core}

\author{Riccardo Fantoni}
\email{riccardo.fantoni@scuola.istruzione.it}
\affiliation{Universit\`a di Trieste, Dipartimento di Fisica, strada
  Costiera 11, 34151 Grignano (Trieste), Italy}

\date{\today}

\begin{abstract}
We performed numerical (restricted) path integral Monte Carlo experiments
on metallic Hydrogen from first principles. We study a quantum two 
component plasma where one component is made of pointwise particles of 
negative unitary charge and the other is made of charged hard spheres of 
positive unitary charge. We study both the additive mixture and a 
nonadditive mixture where we only keep a hard core between unlike species.
We specialize to the case of the electron-proton plasma with a 
1:1 ratios between the molar fraction of the two species. We measured 
thermodynamic and structural properties of the plasma. From an analysis
of the structure we see a transition from a metallic Hydrogen phase, to
a molecular Hydrogen phase as the temperature is lowered. As expected at 
high density the correlations are diminished.
\end{abstract}

%\pacs{...}

\maketitle
\tableofcontents
%%%%%%%%%%%%%%%%%%%%%%%%%%%%%%%%%%%%%%%%%%%%%%%%%%%%%%%%%%%%%%%%%%%%%%%%%%%%%%
\section{Introduction}
%%%%%%%%%%%%%%%%%%%%%%%%%%%%%%%%%%%%%%%%%%%%%%%%%%%%%%%%%%%%%%%%%%%%%%%%%%%%%%
\label{sec:intro}

Hydrogen is the simplest element in the Universe. It has a nucleus made of 
only one proton, the ion, and an orbiting electron. An interesting question is
whether the laws of quantum statistical physics and of Coulomb are able to 
reproduce the 
{\sl formation of the Hydrogen atoms} given a plasma made of an equal number of
electrons and protons, a Two Component Plasma (TCP). This problem has been
widely studied in literature starting from the pioneering exact analytic
study of the non quantum TCP in one dimension of Edwards and Lenard 
\cite{Lenard1961,Edwards62} to reach the realistic first principles ground state 
Monte Carlo calculations of Ref. \cite{Ceperley1981} and non zero temperature Path 
Integral Monte Carlo (PIMC) calculations of Refs. \cite{Pierleoni1994,Bonitz2024} 
or the coupled electron-ion Monte Carlo \cite{Pierleoni2004} where the 
ground state properties of the electrons component are used to perform
a PIMC calculation on the protons component only.

There are many non quantum simulations of the TCP studying the clustering 
and phase transition properties \cite{Fantoni13f}. Here we want to study 
the quantum version specializing to the electron-proton binary mixture with
a 1:1 ratios between the particle numbers of the two species. This will 
allow us to investigate the Hydrogen atom formation problem.

We then start from a completely ionized TCP with a number $N_e=N/2$ 
of polarized electrons equal to the number $N_p=N/2$ of polarized 
protons in a volume $\Omega$ and in thermal equilibrium 
at an inverse temperature $\beta=1/k_BT$ with $k_B$ Boltzmann constant 
and $T$ the absolute temperature. We use a PIMC calculation 
\cite{Ceperley1995} with a number $M=\beta/\tau$ of timeslices 
of length $\tau$ which discretize the imaginary time interval $[0,\beta[$
to determine the thermodynamic and structural properties of the plasma.
In particular the structural clustering that we observe proves the 
ability of the quantum statistical physics theory to predict the 
atom formation from first principles when one introduces just 
Coulomb pair interactions.
\footnote{It would be interesting to see if, for more complicated atoms, 
the introduction of nuclear forces would still make it possible to
predict the atom formation.}

Since the proton is about 2000 times more heavy than the electron we
expect the protons species paths to be much more contracted than the
ones of the electrons species. But, as we will see, this occurrence does 
not pose any serious limitation to what can be simulated but only an 
increase of the computer time needed to measure the properties of
the protons component.

We know that both the electron and the proton are spin $1/2$ 
identical particles and as such they each obey to the Fermi-Dirac 
statistics. But as we will see there is just a slight difference between 
treating both species as fermions or as bosons. This is essentially due 
to the property of bosons to `like' themselves and of the fermions 
to `dislike' themselves. While bosons simulations are 
exact, fermions simulations are affected by the so called `fermions 
sign problem' \cite{Ceperley1991} which can be overcome only by 
approximate methods like the Restricted PIMC (RPIMC) \cite{Ceperley1996}.
The approximate RPIMC for fermions is much more slow than the exact PIMC
for bosons since it requires the calculation of $M$ determinants of 
dimension $N/2$ at each Metropolis attempted move \cite{Kalos-Whitlock}.
So that only small $N$ systems can be studied.

Due to the divergence of the Coulomb potential at contact like $1/r$
it is necessary to introduce a hard core of diameter $\sigma$ for the 
nuclei, the protons, 
while we will consider the electrons as dimensionless and pointwise.
This is needed in order to obtain a stable many body system that does
not collapse. As we will see, for a given temperature $T$, the minimum
hard core diameter preventing the collapse will be dependent from 
temperature, $\sigma=\sigma_{\mbox{min}}(T)$. We will discuss that 
the artificial cutoff $\sigma$ is only necessary in a primitive 
approximation to the action \cite{Ceperley1995}, where the bare 
Coulomb potential directly enters the inter-action. The cutoff
is not necessary anymore in a pair-product approximation to the 
action (see section IV.F of Ref. \cite{Ceperley1995}) that we will 
treat in a forthcoming work.

We will investigate both the additive mixture when the pointwise 
electrons cannot get closer than $\sigma/2$ to the protons
and two protons cannot get closer than $\sigma$ among themselves
and the nonadditive mixture \cite{Fantoni04a,Fantoni13f} when both 
the electrons and the protons are pointwise among themselves but 
the electron still cannot get closer than $\sigma/2$ to a proton.
This analysis will show how introducing a hard core between the 
protons is largely unnecessary. In fact the nonadditive scenario is 
sufficient to ``dress'' the protons with the electronic cloud which
is able to keep two protons roughly a distance $\sigma$ apart 
spontaneously.

We expect a phase transition from a classical TCP at high temperature, 
$T>10^4\mbox{K}$, to
metallic Hydrogen, $10^3\mbox{K}<T<10^4\mbox{K}$, to molecular Hydrogen, 
$10\mbox{K}<T<10^3\mbox{K}$, to solid Hydrogen at lower temperatures 
(see figure 4 of Ref. \cite{Bonitz2024}). A liquid-liquid phase transition 
at approximately $500\mbox{GPa}$ separates the molecular Hydrogen 
at lower pressures to the atomic Hydrogen at higher pressures. In this 
work we will use the canonical $N,n,T$ ensemble where $n$ is the number
density and will simulate just the two metallic and molecular phases.

%%%%%%%%%%%%%%%%%%%%%%%%%%%%%%%%%%%%%%%%%%%%%%%%%%%%%%%%%%%%%%%%%%%%%%%%%%%%%%
\section{Numerical results}
%%%%%%%%%%%%%%%%%%%%%%%%%%%%%%%%%%%%%%%%%%%%%%%%%%%%%%%%%%%%%%%%%%%%%%%%%%%%%%
\label{sec:H,H2,H3}

Let us consider an additive mixture of polarized electrons, $e$, and 
polarized protons, $p$. 
\footnote{Here we choose polarized particles for simplicity but in the
most general setting one would require a four component mixture with
spin up and down electrons and spin up and down protons.}
First of all let us define our reduced units. We measure masses in units of the 
electron mass $m_e$ so that the reduced mass of the electron will be
$\mu_e=m_e/m_e=1$ and the reduced mass of the proton 
$\mu_p=m_p/m_e\approx 1836.15$. The characteristic kinetic energy is
\bq
\calk=\frac{\hbar^2}{2m_e\ell^2},
\eq
with $\ell$ a length. The characteristic Coulomb energy is
\bq
\calv=\frac{e^2}{\ell}=\frac{2\calk\ell}{\call},
\eq
with $e$ the electron charge magnitude and 
\bq \label{eq:Elle}
\call\approx 5.29177\times 10^{-9}\mbox{cm}
\eq
the Bohr radius $a_B=\hbar^2/m_ee^2$. Then, if we denote with $n=N/\Omega$ 
the reduced number density this will correspond to a real number density of 
$n\call^{-3}$ in cm$^{-3}$. For example for metallic Hydrogen
we have a mass density $\approx 1\mbox{g}/\mbox{cm}^3$ which corresponds to a 
reduced number density of $n\approx\call^32/m_p\approx 0.177189$. 

We may then take as characteristic temperature
\bq
\calt=\frac{2\calk\ell^2}{\call^2k_B}\approx 315775\mbox{K}.
\eq
So that for metallic Hydrogen in planetary interiors, like for Jupiter for 
example, an absolute temperature of $\approx 10^4\mbox{K}$ will correspond 
to a reduced temperature of $T\approx 10^4/\calt\approx 0.0316681$. 

The diameter of the nucleus of the Hydrogen atom is of the order of a fermi, i.e.
$\approx 1\mbox{fm}=1\times 10^{-13}\mbox{cm}$ which in reduced units is 
$\sigma_{\mbox{H}}\approx 1.88973\times 10^{-5}$. On the other hand we 
will consider the electrons as pointwise. The electrons will collapse 
onto the protons as soon as their kinetic energy is less than the Coulomb 
attraction at contact. In reduced units this balance can be rudely estimated 
by $T \lesssim 1/\sigma$. 
Therefore choosing the Hydrogen nucleus size, $\sigma_{\mbox{H}}$, as the 
protons hard core, $\sigma$, requires a temperature larger than 
$10^{10}\mbox{K}$ to grant ionization of the mixture against collapse. 
Moreover, upon collapse, the Coulomb energy at contact between a proton
and the collapsed electron will be $1/\sigma_{\mbox{H}}$ in units of 
$e^2/a_B$, i.e. of two Rydbergs. If we have collapse and we use Bose-Einstein 
statistics between the like particles the whole system of (neutral) atoms may 
further collapse to a point, a giant cluster of atoms.

In order to estimate the cutoff reduced distance $r_0=\sigma/2$ in the
Coulomb attractive potential between an electron and a proton we note that
the path averaging in the Feynman-Kac formula smooths the potential. The 
smoothing makes the action finite at the origin instead of having a 
$1/r$ singularity \cite{Ceperley1995,Pollock1988}. The cusp condition for
the pair Coulomb density matrix requires a behavior as 
$\sim r/(1/\mu_1+1/\mu_2)\sim r\mu_2$ as the electron approaches a proton
at a small distance $r$. Setting this equal to their primitive  
inter-action $\tau/r$ we find
\bq \label{eq:r0}
r_0\sim \sqrt{\frac{\tau}{\mu_2}}\sim 0.023\sqrt{\tau},
\eq
in reduced units.

Summarizing, in our reduced units we will measure lengths in units of a Bohr
radius $a_B$ and energies in units of two Rydbergs $e^2/a_B$, i.e. of a Hartree.
The Hamiltonian and the density matrix in reduced units will then read
\bq
H&=&K+V=-\sum_{\alpha=1}^N\frac{1}{2\mu_{i_\alpha}}\nabla^2_\alpha+
\sum_{\alpha<\beta=1}^N\phi_{i_\alpha i_\beta}(r_{\alpha\beta}),\\ \label{eq:dm}
\rho&=&e^{-H/T}.
\eq
where we indicate with a Greek index the particle label and with a Roman index 
the species label so that for example $i_\alpha$ is the species label of 
particle $\alpha$, either 1 for the electrons or 2 for the protons, 
$\rr_\alpha$ is the position vector of particle $\alpha$, and 
$r_{\alpha\beta}=|\rr_{\alpha\beta}|=|\rr_\alpha-\rr_\beta|$ is the 
distance between particles $\alpha$ and $\beta$.
Distances are measured in units of $\call$.
The reduced masses are then $\mu_1=\mu_e$ and $\mu_2=\mu_p$. The partial 
pair potential $\phi_{ij}$ is the usual Coulomb potential given by 
\bq \label{eq:pp}
\phi_{ij}(r)=\left\{\begin{array}{ll}
\epsilon_{ij}/r & r>\Sigma_{ij}\\
\infty & \mbox{else}
\end{array}\right.,
\eq
where in the Two-Component-Plasma (TCP) we choose
\bq \label{eq:eps}
{\boldsymbol\epsilon}&=&\left(\begin{array}{cc}
1 & -1\\
-1 & 1
\end{array}\right),\\ \label{eq:sig}
{\boldsymbol\Sigma}&=&\left(\begin{array}{cc}
0 & \sigma/2\\
\sigma/2 & \sigma
\end{array}\right).
\eq
Note that at a given $T$ if we choose $\sigma$ too small the whole
system collapses with all the negative light particles falling onto the
heavy positive particles.

We will indicate with $N_i$ the number of particles of species $i=1,2$ and 
$N=\sum_i N_i$ the total number of particles with $x_i=N_i/N$ the molar 
fraction for species $i$ such that $\sum_ix_i=1$. The reduced partial 
number densities $n_i=nx_i$
\footnote{Sometimes it can be useful to work in terms of the 
Wigner-Seitz radius $r_s$ for the electron component, i.e.
$r_s=(4\pi n_1/3)^{-1/3}$ the reduced mean inter-electron distance.}
with $n=N/\Omega$ and, in a cubic box,  
$\Omega=L^3$, with $L$ the reduced length of the box side. 

Note that one could introduce for each species some length scales 
\bq
\calw_i=n_i^{-1/3},~~~i=1,2
\eq
and rescale each reduced coordinate as 
$\tilde{\rr}_\alpha=\rr_\alpha/\calw_{i_\alpha}$. Then the rescaled 
kinetic energy of species $i$, $\tilde{K}_i=\calw_i^2K_i$ and 
the rescaled potential energy of species $i$, 
$\tilde{V}_i=\calw_iV_i$. We then see that as 
$n_i$ increases $V_i$ is penalized respect to $K_i$
and that component tends to behave more like a ``free'' gas
with less correlations. This is confirmed by our numerical 
experiments. 

The box can be made periodic so that it permeates the whole infinite three
dimensional space in order to mimic the thermodynamic limit. This will
produce finite size errors that can be reduced by increasing the box size 
$L$. Of course in order to study a given thermodynamic state of the system 
this requires increasing $L$ at constant density $n$ which means increasing 
$N$. This will make the computer experiment more demanding from the 
numerical calculation cost point of view. 

Moreover, in periodic boundary conditions the potential energy $V$ for 
the particles within the central box should include also the pair 
interactions with all their images in the periodic boxes.
Since the Coulomb potential $1/r$ is long range these contributions 
cannot generally be neglected and specific techniques should be used 
\cite{Natoli1995} to treat them, the most common of which is the Ewald 
summation. In this preliminary work we will simply neglect 
interactions with periodic images since we will not carry out a
careful finite size error analysis.

We performed some computer experiments as described in Table \ref{tab:tq}. 
During the simulations we measured some thermodynamic quantities as the 
total kinetic and potential energy of the TCP and the superfluid
fractions for each species. These were measured using the thermodynamic
estimators described in Ref. \cite{Ceperley1995}. And we also calculated 
the partial and total radial distribution functions \cite{Allen-Tildesley}. 
Given an observable 
$\calo$ we can measure it during the simulation through the following 
PIMC operation \cite{Ceperley1995} 
\bq
\langle\calo\rangle &=&\mbox{tr}(\rho\calo)/Z,\\
Z&=&\mbox{tr}(\rho),
\eq
where $\rho$ is the primitive approximation \cite{Ceperley1995} for the
density matrix of Eq. (\ref{eq:dm}), $\mbox{tr}(\ldots)$
is the trace operation, and $Z$ is the canonical partition function. 
Therefore we impose both spatial and imaginary time periodic boundary 
conditions for the many body path of both species 
$R(t)=(\rr_1(t),\rr_2(t),\ldots,\rr_N(t))$ where $t$ is the imaginary 
time. So that for any function $f$ 
we require $f(R+L\uu)=f(R)$ with $\uu$ a unit vector along any of the 
$3N$ dimensions and $R(t+\beta)=R(t)$ with $\beta=1/T$
the reduced inverse temperature. Our algorithm is discussed in the
Appendix \ref{app:alg}

\begin{table*}[htbp]
\caption{Cases treated in our simulations for the additive TCP 
  with a pair Coulomb potential as in Eq. (\ref{eq:pp}) with (\ref{eq:eps}) 
  and (\ref{eq:sig}) (cases 
  C,D,E,F,G correspond to the electron-proton plasma). The 
  reduced masses of the two species are $\mu_1=1$ and $\mu_2$, the
  molar fractions $x_1=1-x_2$ and $x_2$, $\sigma$ the hard core of species 2,
  the total number of particles $N$, the reduced density $n=N/\Omega$, 
  the reduced inverse temperature 
  $\beta=1/T$, $e_K=\langle K\rangle/N$ the total kinetic energy per particle,
  $e_V=\langle V\rangle/N$ the total potential energy per particle, $f_s(i)$
  the superfluid fraction for species $i$. These three thermodynamic 
  quantities are measured from the thermodynamic estimators discussed 
  in Ref. \cite{Ceperley1995}. The other quantities were introduced in 
  the main text. 
  In the statistics column ``b'' stands for bosons, ``f'' for fermions. 
  The simulation lasted more than $10^8$ Monte Carlo steps
  where one step is made of a displace move of all the beads of a single
  particle path and a bridge move. The simulations for fermions lasted
  about one week of computer time.}  
\label{tab:tq}
\begin{ruledtabular}
\begin{tabular}{|c|llllllllllll|}
\hline
case & statistics & $\beta$& $M$ & $n$ & $N$ & $x_2$ & $\mu_2$ & $\sigma$ & $Ne_K$ & $-Ne_V$ & $f_s(1)$ & $f_s(2)$ \\ 
\hline
A & b (PIMC) & 10 & 20 & 0.1 & 20 & 0.5 & 1000 & 0.4 & 5.26(2) & 11.95(4) & 0.520(5) & 0.0477(2)\\
B & f (RPIMC)& 10 & 20 & 0.1 & 20 & 0.5 & 1000 & 0.4 & 4.47(7)  & 12.21(6)  & 0.44(3) & 0.0474(2)\\
C & b (PIMC) & 10 & 20 & 0.08 & 20 & 0.5 & 1836.15 & 0.4 & 5.54(2) & 12.81(4) & 0.370(6) & 0.0477(6)\\
D & f (RPIMC)& 10 & 20 & 0.08 & 20 & 0.5 & 1836.15 & 0.4 & 4.17(7)  & 11.25(5)  & 0.62(5) & 0.0476(3)\\
E & b (PIMC) & 10 & 20 & 0.18 & 20 & 0.5 & 1836.15 & 0.4 & 5.86(2) & 13.38(3) & 0.557(5) & 0.0478(2)\\
F & f (RPIMC)& 10 & 20 & 0.18 & 20 & 0.5 & 1836.15 & 0.4 & 4.25(7) & 12.98(5) & 0.71(5) & 0.0477(3)\\
G & b (PIMC) & 200 & 20 & 0.16 & 20 & 0.5 & 1836.15 & 1 & 1.709(1) & 23.56(2) & 0.093(4) & 0.0469(4)\\
H & f (RPIMC)& 200 & 20 & 0.16 & 20 & 0.5 & 1836.15 & 1 & 1.713(1) & 24.283(5) & 0.230(5) & 0.0469(7)\\
\hline
\end{tabular}
\end{ruledtabular}
\end{table*}
\begin{figure}[htbp]
\begin{center}
\includegraphics[width=8cm]{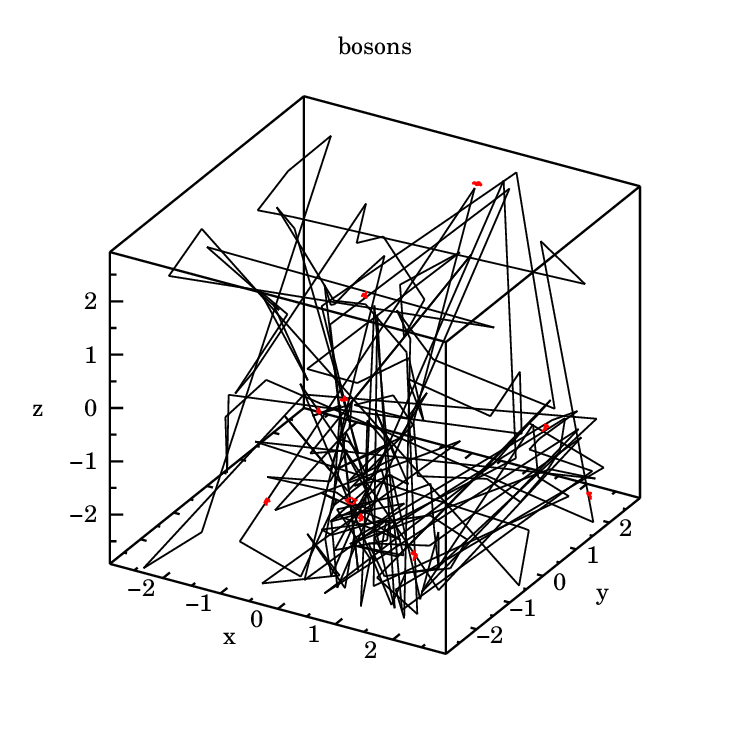}
\includegraphics[width=8cm]{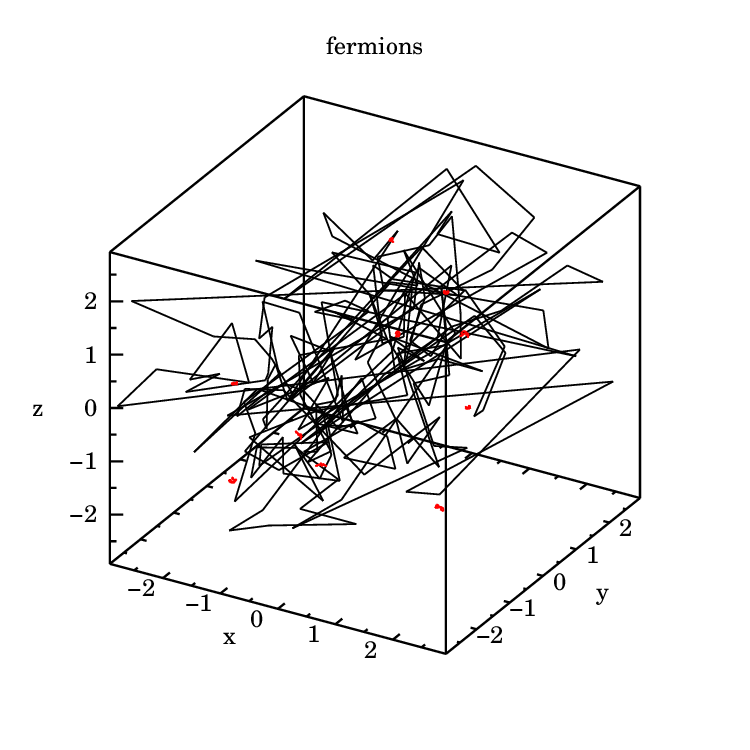}
\end{center}  
\caption{Snapshots of the simulation of a binary mixture with 
$N=20, x_1=1/2, \mu_1=1,\mu_2=1000,\sigma=0.4$. The mixture 
is in a thermodynamic state with a reduced temperature $T=T'/\calt=0.1$ 
with $T'$ measured in degrees Kelvin and a reduced number density 
$n=n'\call^3=0.1$ with $n'$ measured in cm$^{-3}$. 
In the path integral we chose $M=20$ and used either 
Bose-Einstein (left panel, case A in Table \ref{tab:tq}) 
or Fermi-Dirac (right panel, case B in Table \ref{tab:tq}) statistics 
among like particles. Red paths are for the heavy slow species and black 
paths are for the light fast species.} 
\label{fig:snap}
\end{figure}
\begin{figure}[htbp]
\begin{center}
\includegraphics[width=8cm]{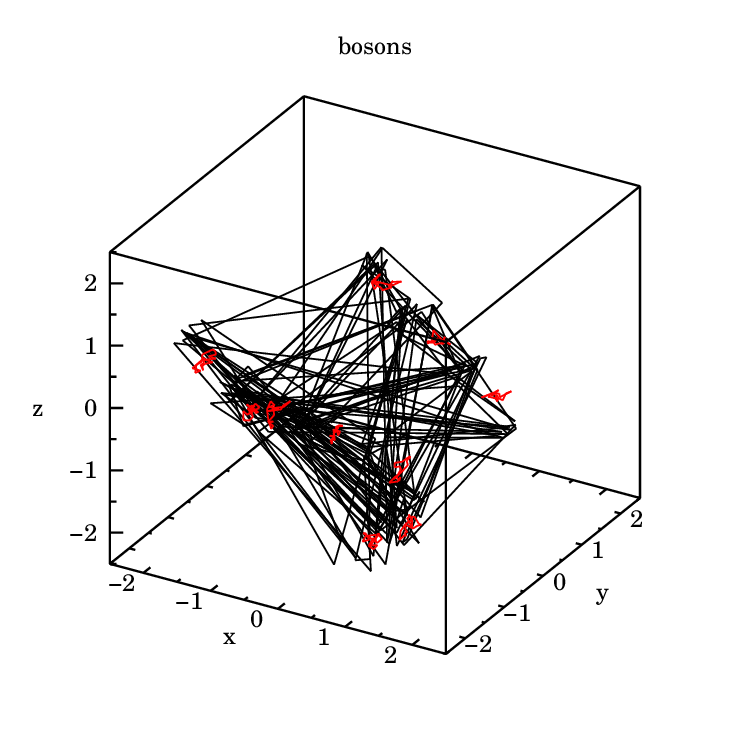}
\includegraphics[width=8cm]{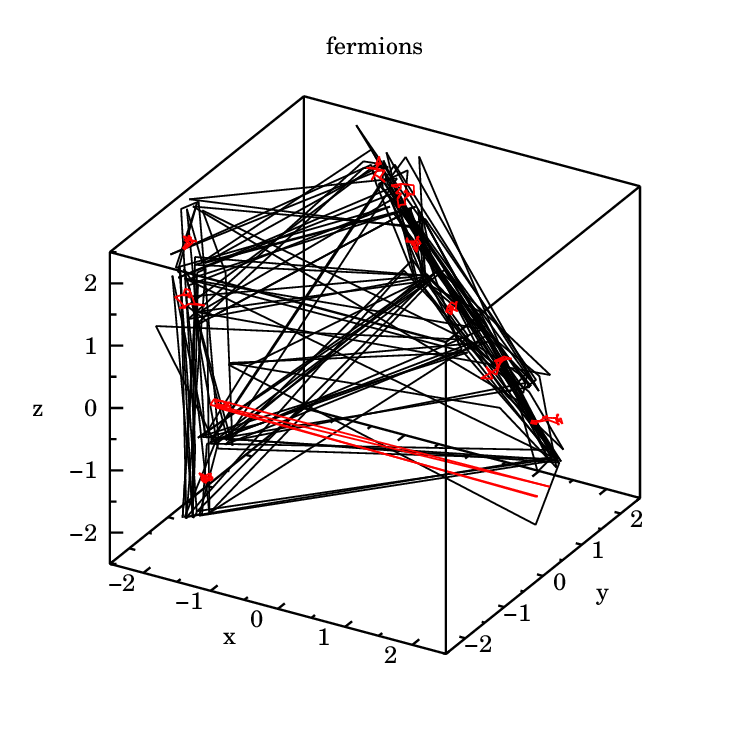}
\end{center}  
\caption{Snapshots of the simulation of a binary mixture with 
$N=20, x_1=1/2, \mu_1=1,\mu_2=1836.15,\sigma=1$. The mixture 
is in a thermodynamic state with a reduced temperature $T=T'/\calt=0.005$ 
with $T'$ measured in degrees Kelvin and a reduced number density 
$n=n'\call^3=0.16$ with $n'$ measured in cm$^{-3}$. 
In the path integral we chose $M=20$ and used either 
Bose-Einstein (left panel, case G in Table \ref{tab:tq}) 
or Fermi-Dirac (right panel, case H in Table \ref{tab:tq}) statistics 
among like particles. Red paths are for the heavy slow protons and black 
paths are for the light fast electrons.} 
\label{fig:snapH2}
\end{figure}
\begin{figure}[htbp]
\begin{center}
\includegraphics[width=8cm]{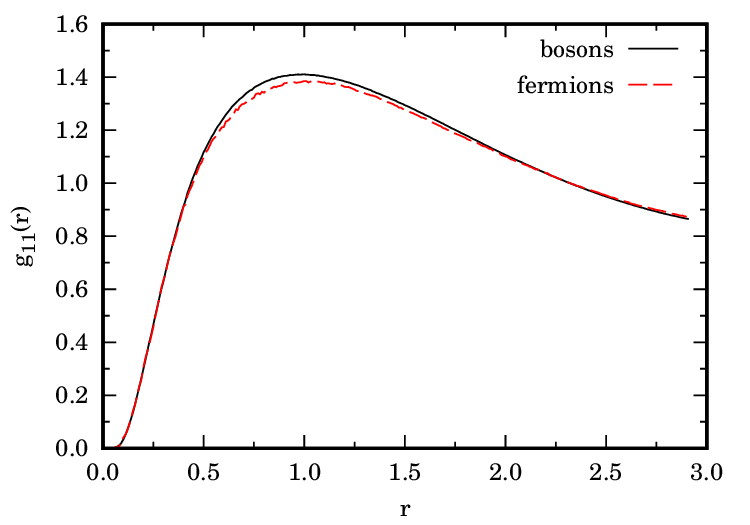}
\includegraphics[width=8cm]{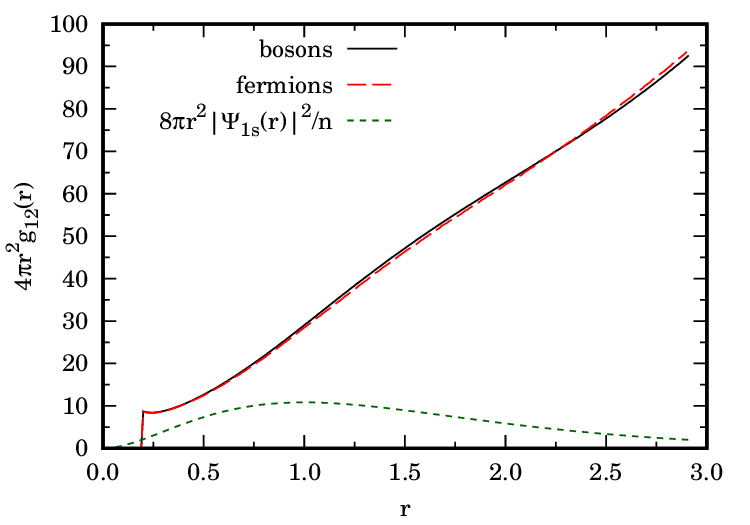}\\
\includegraphics[width=8cm]{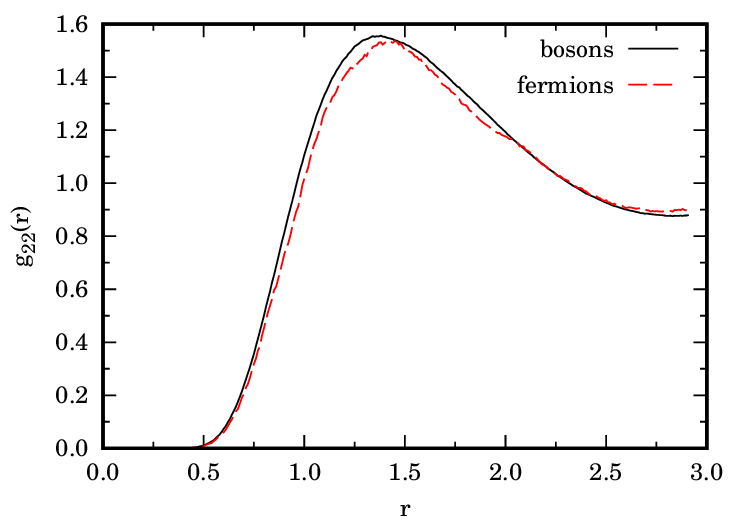}
\includegraphics[width=8cm]{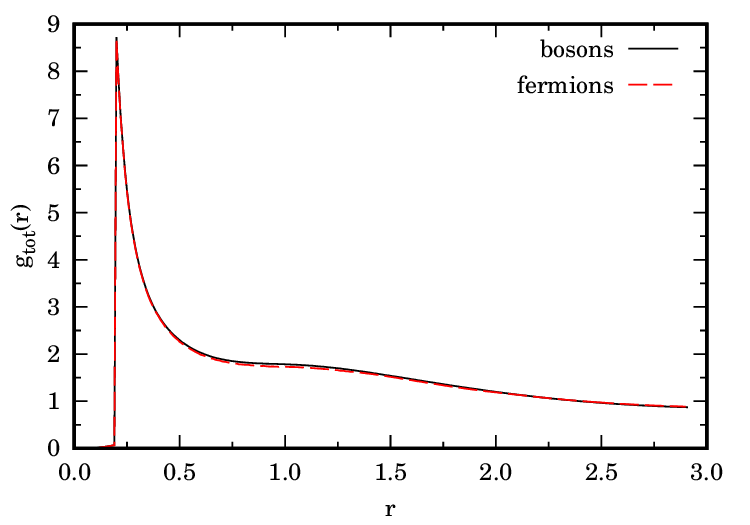}
\end{center}  
\caption{The partial, $g_{ij}$, and total, $g_{tot}$, radial distribution 
functions for a binary mixture with 
$N=20, x_1=1/2, \mu_1=1,\mu_2=1000,\sigma=0.4$. In the panel for the 
unlike radial distribution function we also show the ground state 
probability distribution of the Hydrogen atom for comparison. The mixture 
is in a thermodynamic state with a reduced temperature $T=T'/\calt=0.1$ 
with $T'$ measured in degrees Kelvin and a reduced number density 
$n=n'\call^3=0.1$ with $n'$ measured in cm$^{-3}$. 
In the path integral we chose $M=20$ and used either 
Bose-Einstein (case A in Table \ref{tab:tq}) or 
Fermi-Dirac (case B in Table \ref{tab:tq}) statistics among like particles. 
In the figures $r=r'/\call$ where $r'$ is measured in centimeters.} 
\label{fig:gij}
\end{figure}
\begin{figure}[htbp]
\begin{center}
\includegraphics[width=8cm]{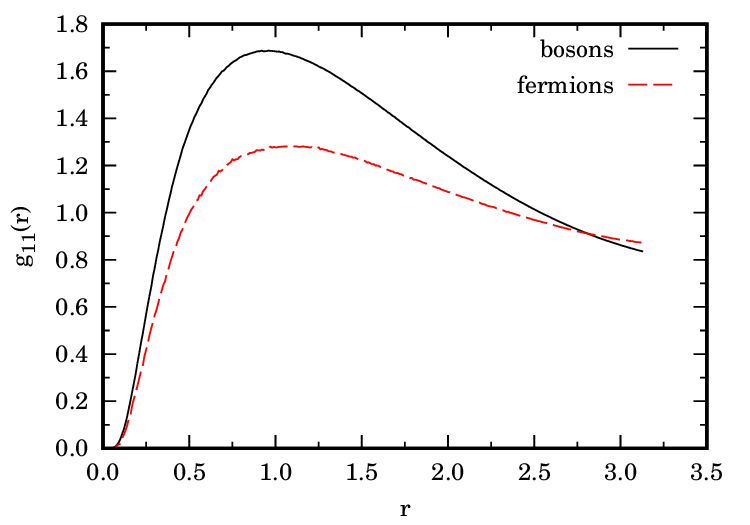}
\includegraphics[width=8cm]{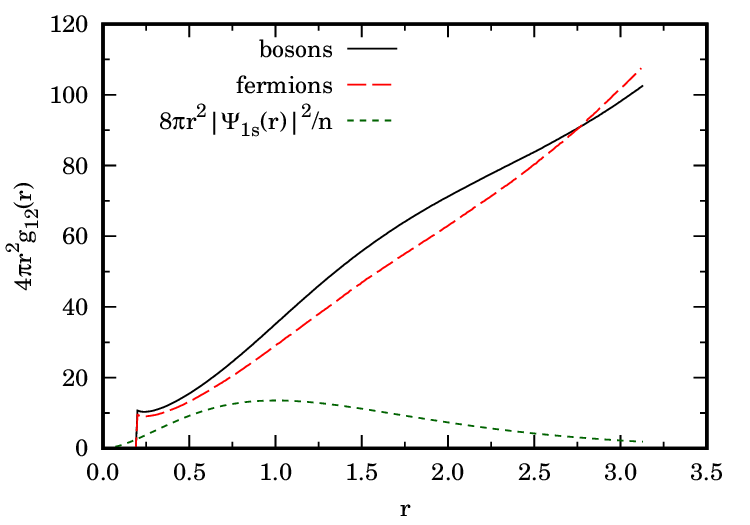}\\
\includegraphics[width=8cm]{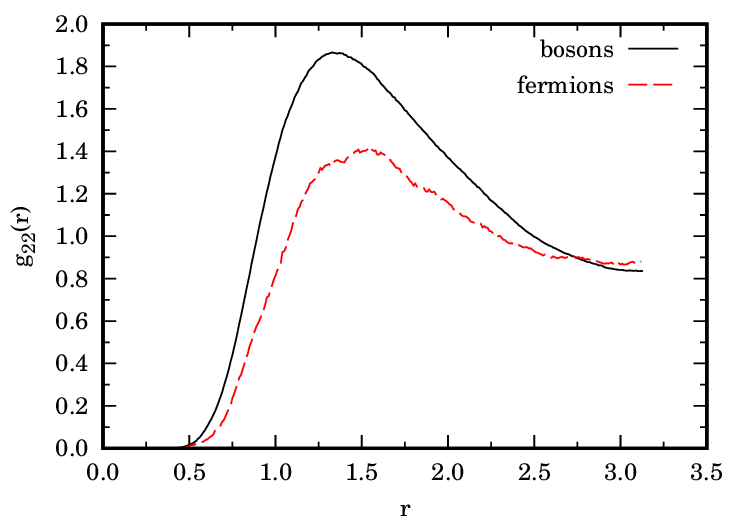}
\includegraphics[width=8cm]{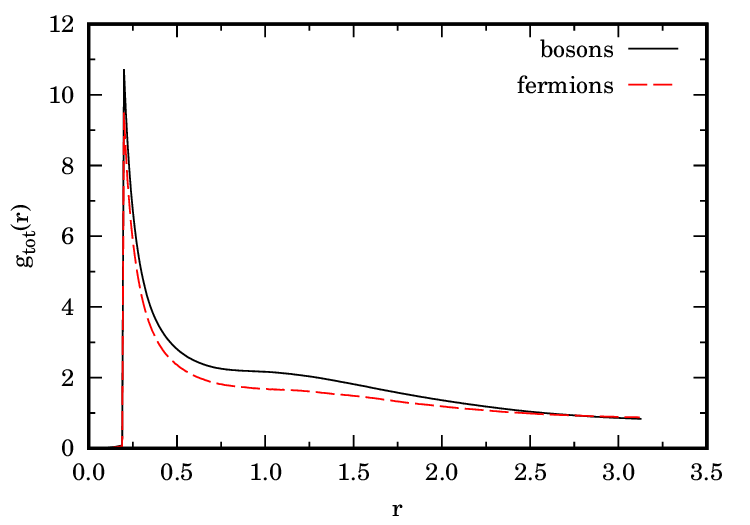}
\end{center}  
\caption{The partial, $g_{ij}$, and total, $g_{tot}$, radial distribution 
functions for a binary mixture with 
$N=20, x_1=1/2, \mu_1=1,\mu_2=1836.15,\sigma=0.4$. In the panel for the 
unlike radial distribution function we also show the ground state 
probability distribution of the Hydrogen atom for comparison. The mixture 
is in a thermodynamic state with a reduced temperature $T=T'/\calt=0.1$ 
with $T'$ measured in degrees Kelvin and a reduced number density 
$n=n'\call^3=0.08$ with $n'$ measured in cm$^{-3}$. 
In the path integral we chose $M=20$ and used either 
Bose-Einstein (case C in Table \ref{tab:tq}) or 
Fermi-Dirac (case D in Table \ref{tab:tq}) statistics among like particles. 
In the figures $r=r'/\call$ where $r'$ is measured in centimeters.} 
\label{fig:gijH}
\end{figure}
\begin{figure}[htbp]
\begin{center}
\includegraphics[width=8cm]{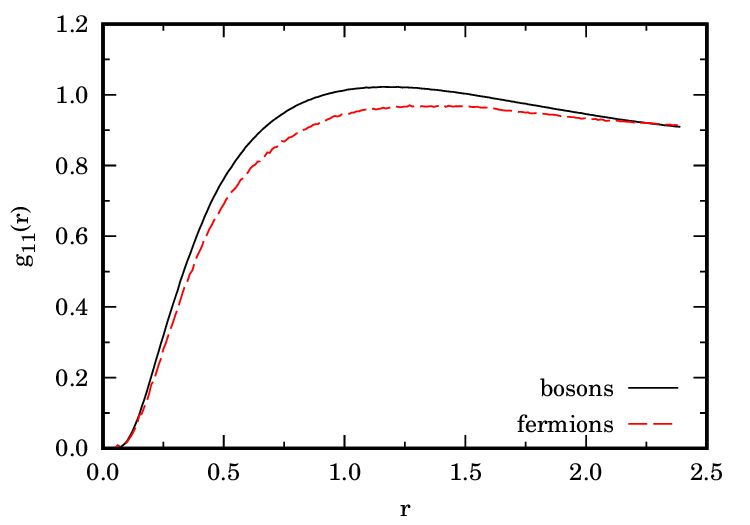}
\includegraphics[width=8cm]{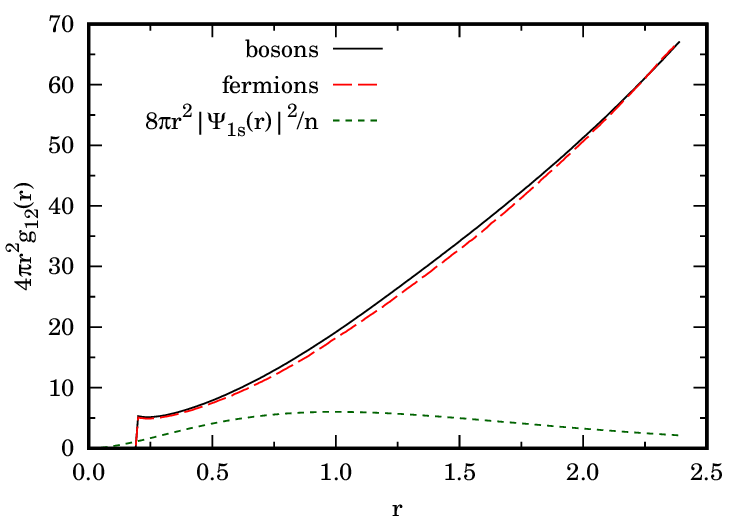}\\
\includegraphics[width=8cm]{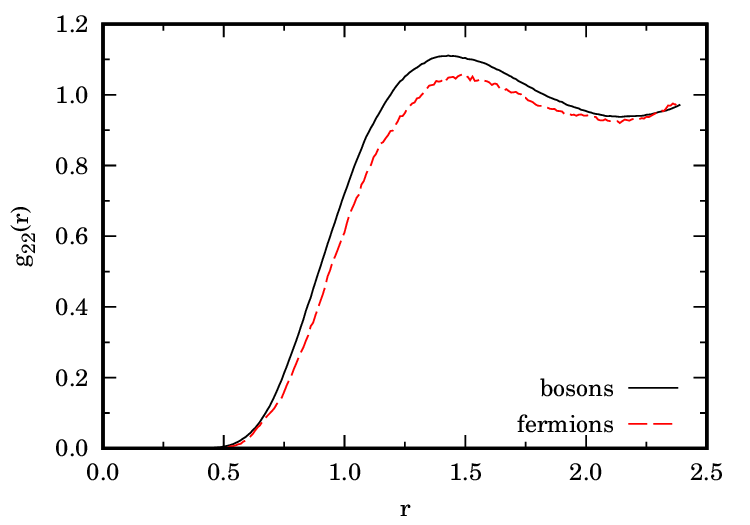}
\includegraphics[width=8cm]{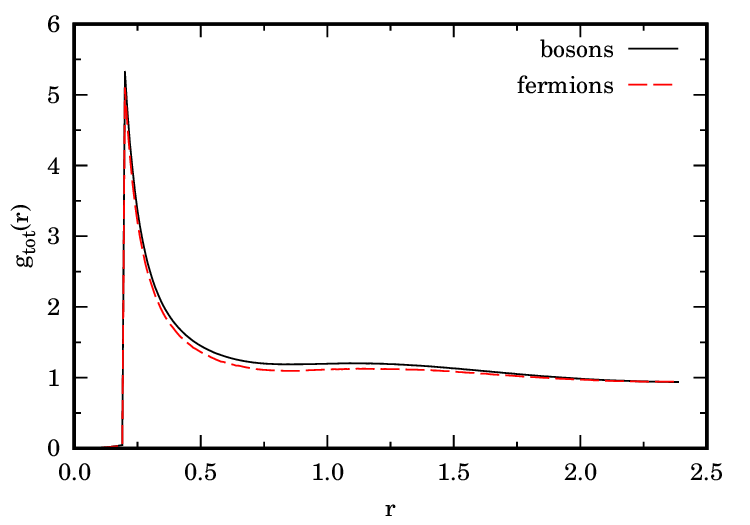}
\end{center}  
\caption{The partial, $g_{ij}$, and total, $g_{tot}$, radial distribution 
functions for a binary mixture with 
$N=20, x_1=1/2, \mu_1=1,\mu_2=1836.15,\sigma=0.4$. In the panel for the 
unlike radial distribution function we also show the ground state 
probability distribution of the Hydrogen atom for comparison. The mixture 
is in a thermodynamic state with a reduced temperature $T=T'/\calt=0.1$ 
with $T'$ measured in degrees Kelvin and a reduced number density 
$n=n'\call^3=0.18$ with $n'$ measured in cm$^{-3}$. 
In the path integral we chose $M=20$ and used either 
Bose-Einstein (case E in Table \ref{tab:tq}) or 
Fermi-Dirac (case F in Table \ref{tab:tq}) statistics among like particles. 
In the figures $r=r'/\call$ where $r'$ is measured in centimeters.} 
\label{fig:gijH1}
\end{figure}
\begin{figure}[htbp]
\begin{center}
\includegraphics[width=8cm]{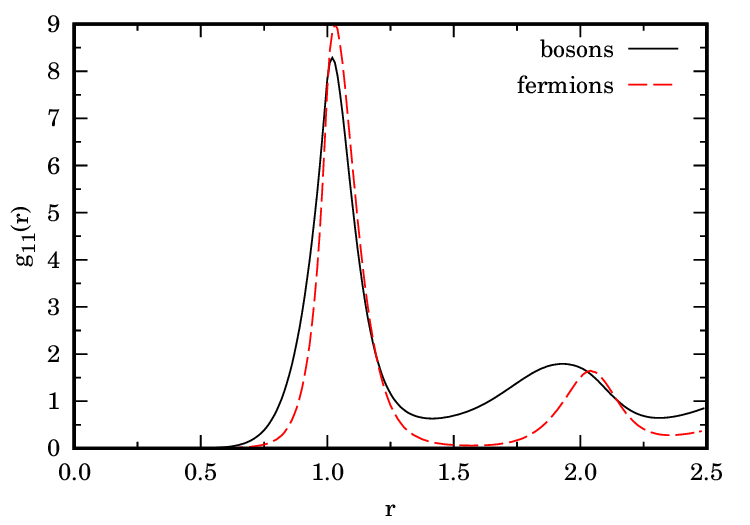}
\includegraphics[width=8cm]{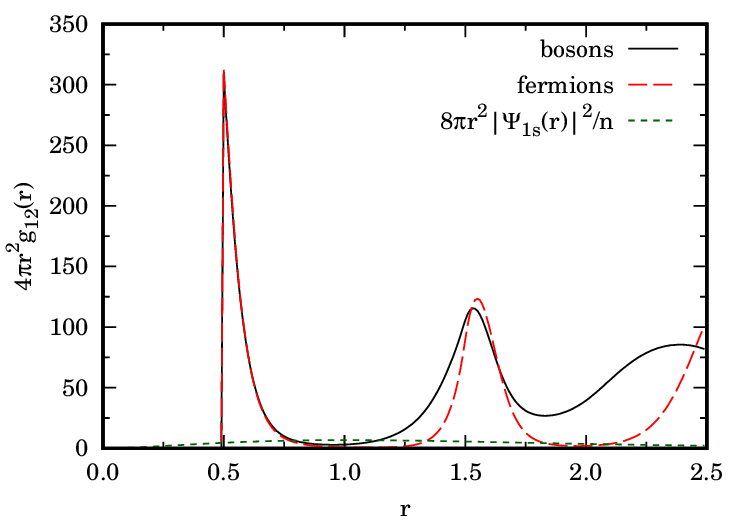}\\
\includegraphics[width=8cm]{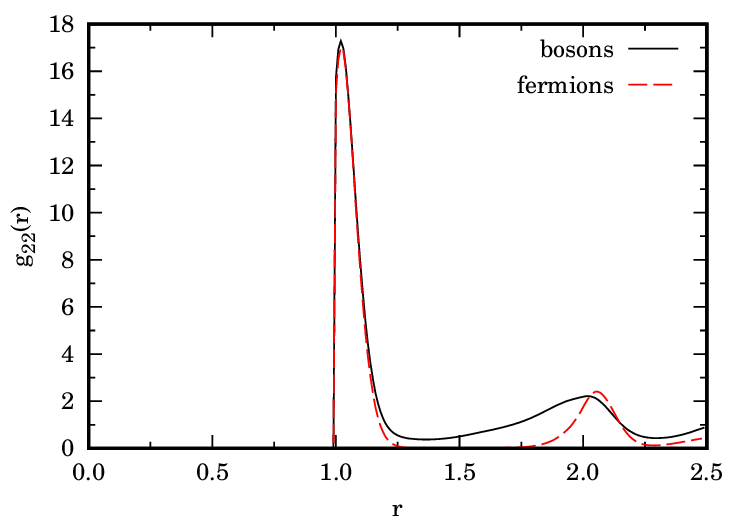}
\includegraphics[width=8cm]{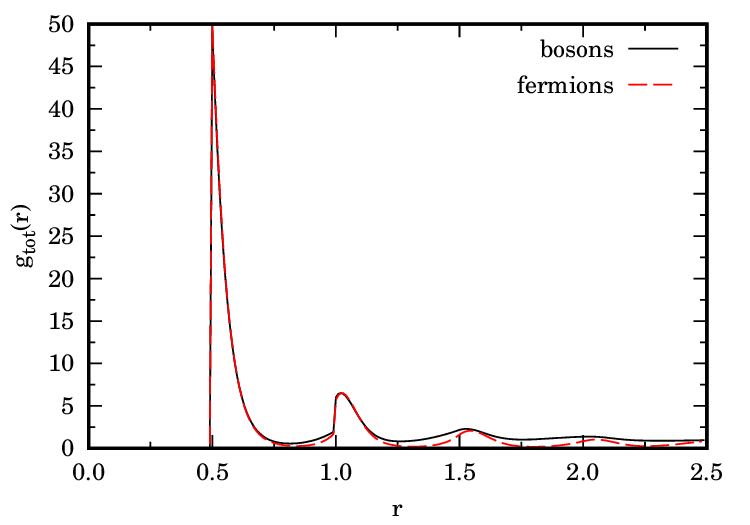}
\end{center}  
\caption{The partial, $g_{ij}$, and total, $g_{tot}$, radial distribution 
functions for a binary mixture with 
$N=20, x_1=1/2, \mu_1=1,\mu_2=1836.15,\sigma=1$. In the panel for the 
unlike radial distribution function we also show the ground state 
probability distribution of the Hydrogen atom for comparison. The mixture 
is in a thermodynamic state with a reduced temperature $T=T'/\calt=0.005$ 
with $T'$ measured in degrees Kelvin and a reduced number density 
$n=n'\call^3=0.16$ with $n'$ measured in cm$^{-3}$. 
In the path integral we chose $M=20$ and used either 
Bose-Einstein (case G in Table \ref{tab:tq}) or 
Fermi-Dirac (case H in Table \ref{tab:tq}) statistics among like particles. 
In the figures $r=r'/\call$ where $r'$ is measured in centimeters.} 
\label{fig:gijH2}
\end{figure}

In Fig. \ref{fig:snap} we show snapshots taken during the
simulations of cases A and B of Table \ref{tab:tq} at equilibrium. 
We see how the light and fast species is in a quantum regime with 
extended paths while the heavy and slow species is in a 
classical regime with contracted paths. In these cases 
$\tau=\beta/M=0.5$ which is small. Fig. \ref{fig:snapH2} 
shows the snapshots for the electron-proton plasma at low 
temperature of cases G and H of Table \ref{tab:tq} at equilibrium. 
We see how in these cases also the protons are in a quantum regime 
with extended paths. We also see the formation of H atoms and 
of H$_2$ molecules. Note that in these cases $\tau=\beta/M=10$
which is rather big. We did not perform a careful study of the 
imaginary time discretization error since we are here only 
interested in the phenomenological and qualitative properties of 
the simulations. We leave to further works the extrapolation 
to the continuum $\tau\to 0$ limit necessary to extract quantitative 
results. From both Figs. \ref{fig:snap} and \ref{fig:snapH2}
we see how bosons like to stay together while fermions like to
stay apart.

The partial pair correlation functions or partial Radial Distributions 
Functions (RDF), $g_{ij}(r)$, are a sensitive indicator of quantum and 
correlation effects and reflect the structural properties of the TCP. 
The like $i=j$ RDF is proportional to the probability that sitting on 
a particle of species $i$ one has to find another particle of the same 
species within the radial interval $[r,r+dr]$, whereas the 
unlike RDF $i\neq j$ give the structure of species $i$ as seen from 
the other species. The total RDF is defined as
$g_{\rm tot}=x_1^2g_{11}+2x_1x_2g_{12}+x_2^2g_{22}$. 
From Figs. \ref{fig:gij},\ref{fig:gijH},\ref{fig:gijH1}, 
and \ref{fig:gijH2} we see the 
formation of clusters made of an heavy positive particle (of species 2) 
surrounded by a cloud of light negative particles (of species 1). 
In the panel for the unlike RDF we compare with the electronic 
probability density corresponding to the isolated atom ground 
state $|\Psi_{1s}|^2$ which has a characteristic radial length equal to 
a Bohr radius $a_B$. 
In Fig. \ref{fig:gij} we show results for cases A and B
of Table \ref{tab:tq} and we see that the species 1 develops a peak 
in $g_{11}$ at $\sim a_B$ while the species 2 at $\sim 1.3 a_B$
in $g_{22}$. 
In Fig. \ref{fig:gijH} we show results for cases C and D
of Table \ref{tab:tq}, in Fig. \ref{fig:gijH} we show results for 
cases E and F, and In Fig. \ref{fig:gijH} we show results for 
cases G and H. All these cases correspond to a realistic 
electron-proton plasma.
We also see how changing the statistics between (identical) particles 
of the same species from Bose-Einstein to Fermi-Dirac
makes them dislike themselves more as expected from the Pauli exclusion
principle. Comparing Figs. \ref{fig:gijH} and \ref{fig:gijH1} we see how 
the difference between the two statistics increases at lower densities 
for a given low temperature, as expected. Comparing the reduced 
de Broglie thermal wavelength $\Lambda_i=\sqrt{1/T\mu_i}$ with the 
interparticle spacing $n_i^{-1/3}$ we may define a reduced degeneracy
temperature $T^D_i=n_i^{2/3}/\mu_i$ for each species. For temperatures 
higher than the degeneracy temperature, quantum statistics, either
bosonic or fermionic, are not very important. Note that the difference 
between the two statistics for the heavy and slow protons component is 
driven by the light and fast electrons component which ``dress'' the 
protons.

From Fig. \ref{fig:gijH1} we see how the oscillations in the proton-proton 
RDF, $g_{22}$, lower their radial length scale as density is increased, with a 
first peak around $\sim 1.3 a_B$. Moreover the first peak in the like RDFs
gets damped in agreement with our expectation to observe less correlations
in this case. The magnitudes of the total kinetic and potential energies
and of the superfluid fraction of the electron component all increase as
only the number density of the electron-proton plasma is increased. 
As expected the superfluid fraction of the protons component is almost zero
since the protons behave nearly classically without exchanges. So that the
winding number for their paths is almost zero.

In Fig. 16 of Ref. \cite{Bonitz2024} they
found, using the coupled electron-ions PIMC, some evidence for the 
$H_2$ molecule formation at $T=1200\mbox{K}/\calt\approx 3.80017\times 10^{-3}$ 
and $r_s=1.44$ which means $n\approx 0.159902$. 
This is made manifest by the development of a 
pronounced first peak in the proton-proton RDF at these low temperatures. 
In Fig. \ref{fig:gijH2} we present the RDF that we measured close to that 
thermodynamic state. Even if we use a number of particles, $N$, and a 
number of timeslices, $M$, much smaller than the ones used in 
Ref. \cite{Bonitz2024}, we see how our unlike RDFs develops a pronounced 
peak near one Bohr radius, at contact of two hard core ions, as expected by the 
formation of $H_2$ molecules. Moreover we also see the appearance of a 
second peak, much more damped, at two Bohr radii which can be interpreted as 
a trace of the formation of $H_3$ linear molecules. At this low temperature 
also the protons paths begin to show a certain extension so that they cannot
be treated classically anymore. Nonetheless their superfluid fraction is still
very close to zero and their exchanges we observe in the simulation are 
negligible. In these cases, contrary to the other cases treated, also the 
superfluid fraction of the electrons component is almost zero. This is due to
the fact that at this low temperature the electrons do not have enough kinetic 
energy to escape the Coulomb attraction towards the protons. And each electron 
is captured by one proton in a H atom.

\begin{figure}[htbp]
\begin{center}
\includegraphics[width=8cm]{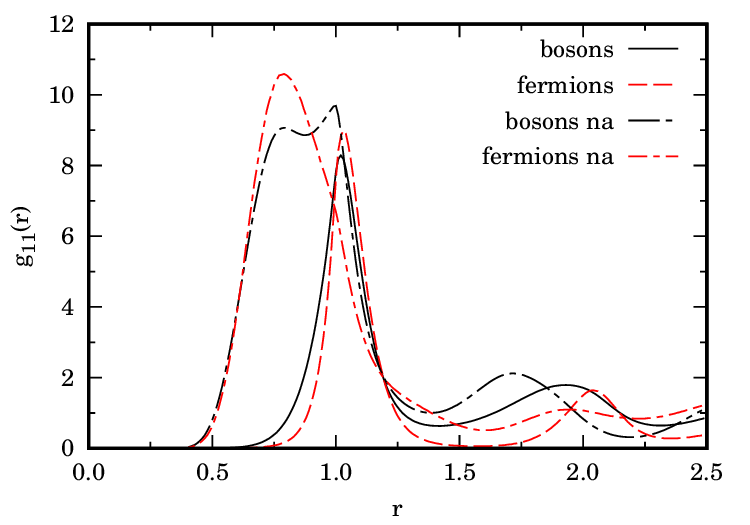}
\includegraphics[width=8cm]{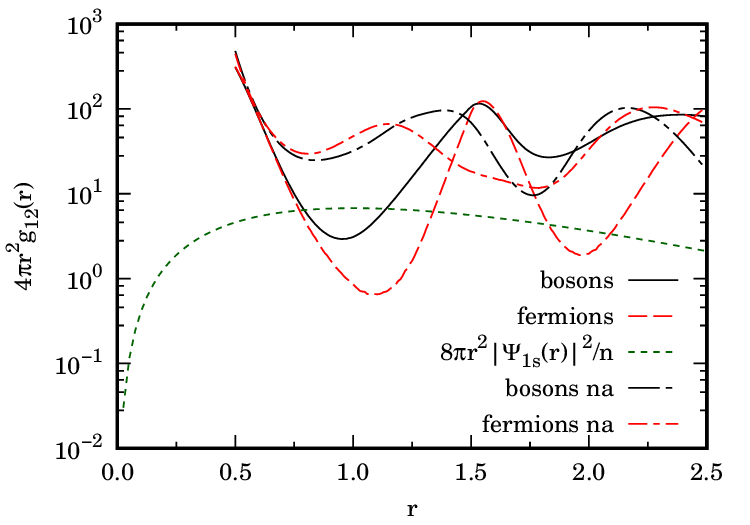}\\
\includegraphics[width=8cm]{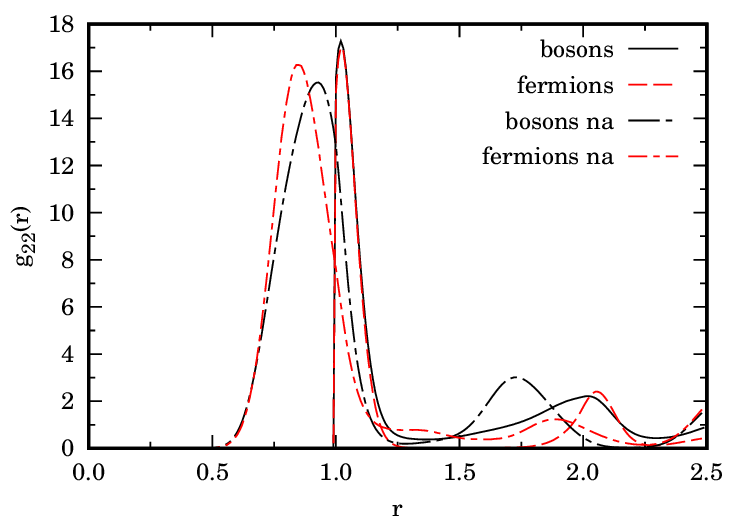}
\includegraphics[width=8cm]{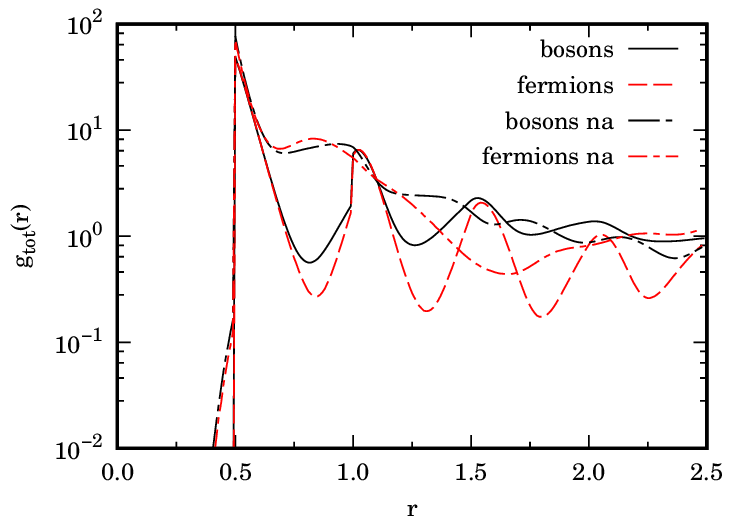}
\end{center}  
\caption{The partial, $g_{ij}$, and total, $g_{tot}$, radial distribution 
functions for the nonadditive (na) binary mixture of Eq. (\ref{eq:signa}) with 
$N=20, x_1=1/2, \mu_1=1,\mu_2=1836.15,\sigma=1$. In the panel for the 
unlike radial distribution function we also show the ground state 
probability distribution of the Hydrogen atom for comparison. The mixture 
is in a thermodynamic state with a reduced temperature $T=T'/\calt=0.005$ 
with $T'$ measured in degrees Kelvin and a reduced number density 
$n=n'\call^3=0.16$ with $n'$ measured in cm$^{-3}$. 
In the path integral we chose $M=20$ and used either 
Bose-Einstein (case G in Table \ref{tab:tq}) or 
Fermi-Dirac (case H in Table \ref{tab:tq}) statistics among like particles. 
In the figures $r=r'/\call$ where $r'$ is measured in centimeters. We also
report the partial RDF of Fig. \ref{fig:gijH2} for comparison. We use a 
logarithmic scale for the unlike and the total RDF.} 
\label{fig:gijH2na}
\end{figure}

Note that instead of the additive mixture of Eq. (\ref{eq:sig}) we could
use a nonadditive one in the Widom-Rowlinson limit 
\cite{Fantoni04a,Fantoni13f} with 
\bq \label{eq:signa}
{\boldsymbol\Sigma}&=&\left(\begin{array}{cc}
0 & \sigma/2\\
\sigma/2 & 0
\end{array}\right).
\eq 
We repeated for this new scenario the simulations of cases G and H. 
In Fig. \ref{fig:gijH2na} we show the partial and total RDF that 
we obtained. From the figure we see how there is a sensible difference 
between the Fermi-Dirac and the Bose-Einstein statistics. For fermions 
the peaks of the partial RDF are shifted at smaller radii
than for bosons. In particular the reduced radius of the H$_2$ 
molecule, the first peak in $g_{22}$, is $\approx 0.95$ in the 
Bose-Einstein case and $\approx 0.85$ in the Fermi-Dirac case. Note 
that even if we did not explicitly impose a hard core among the 
protons nonetheless the first peak in $g_{22}$ approximately
coincides with the additive case with $\sigma=1$. Moreover the
second peak disappears for the fermions case. On the other hand
$g_{11}$ presents a shoulder preceding its first peak in the bosons case
that disappears in the fermions case. Once again the first peak 
in this nonadditive case approximately coincides with the previous
additive case.
%

%%%%%%%%%%%%%%%%%%%%%%%%%%%%%%%%%%%%%%%%%%%%%%%%%%%%%%%%%%%%%%%%%%%%%%%%%%%%%%
\section{Conclusions}
%%%%%%%%%%%%%%%%%%%%%%%%%%%%%%%%%%%%%%%%%%%%%%%%%%%%%%%%%%%%%%%%%%%%%%%%%%%%%%
\label{sec:conclusions}

We simulated with the path integral Monte Carlo method a quantum binary mixture 
where the like particles obeys either to the Bose-Einstein, using an {\sl exact}
 PIMC calculation, or to the Fermi-Dirac statistics, using an
{\sl approximate} RPIMC. We 
used as pair interaction among the particles the Coulomb $1/r$ interaction
assuming total neutrality of the mixture (so to ensure thermodynamic 
stability) and considered the case with repulsion between like particles 
and attraction among unlike particles. This is a Two Component Plasma (TCP).
We then specialized our TCP to the case of the electron-proton plasma 
with 1:1 ratio between the molar fractions of the two species. This allowed
us to investigate the formation of H atoms predicted simply by the laws
of quantum statistical physics and of Coulomb. To stabilize the plasma 
against collapse of the electrons onto the protons we needed to introduce 
a spherical hard core of diameter $\sigma$ for the protons with the electrons 
left pointwise. We considered both the additive and nonadditive cases. 

We then started at high enough temperature, $T$, to ensure a ionized plasma and
gradually lowered it. We first observed the formation of H atoms 
resulting in a metallic Hydrogen phase by looking at the unlike RDF which
develops a strong peak at contact between the pointwise electron and the
hard core ion. We also saw that increasing the density makes the two
components behave more like free gases.
     
Lowering the temperature even more we observed that a phase transition of
the electron-proton plasma from a metallic phase with mostly delocalized 
electrons as in Figs. \ref{fig:gijH} and \ref{fig:gijH1} to a molecular 
fluid phase as in Fig. \ref{fig:gijH2}. Looking at the total RDF $g_{\rm tot}$ 
of Fig. \ref{fig:gijH2} we then see that the first peak is the result
of the H atom formation whereas the second peak reflects the H$_2$ molecule
formation. The molecule formation is made possible by the effect of the
interstitial electrons between two protons which pull the two protons 
one close to the other when the other Coulomb forces from the other
charges around the two protons are screened. There will be a balance 
between the repulsion between the two protons and their attraction due to
the interstitial electrons which determines the first peak in
$g_{22}$.

The same result could have been obtained by keeping the temperature 
constant and lowering the ions hard core diameter $\sigma$. This unphysical
freedom however is only apparent and required by the primitive approximation
\cite{Ceperley1995} which is unbounded from below for the electron-proton 
interaction and therefore breaks down the Trotter approximation 
\cite{Trotter1959} giving rise to the path integral expression for the 
density matrix. Instead of introducing hard cores, an alternative way out 
of this problem, is to truncate the unlike Coulomb potential artificially 
as
\bq \label{eq:ppt}
\phi_{ij}(r)=\left\{\begin{array}{ll}
\epsilon_{ij}/r & r>\Sigma_{ij}\\
\epsilon_{ij}/\Sigma_{ij} & \mbox{else}
\end{array}\right.~~~i\neq j,
\eq
choosing $\sigma=2r_0$ from Eq. (\ref{eq:r0}). Otherwise a route free of 
any artificial parameter is to use the pair Coulomb density matrix
\cite{Ceperley1995,Pollock1988} for the unlike component instead 
of the usual primitive approximation. It is interesting to stress that 
this is only possible in quantum statistical physics and in 
classical statistical physics the hard core between unlike species is 
an unavoidable feature necessary to grant stability of a neutral TCP. 

Regarding the comparison between the simulations using Bose-Einstein or 
Fermi-Dirac statistics between like particles we noticed a sensible difference 
for the nonadditive mixture below the partial degeneracy temperatures. 
The difference still persists for the heavy protons component due to their
interaction with the light electrons component. From Fig. \ref{fig:gijH2na} 
we see how for the nonadditive case in the Widom-Rowlinson 
limit, where the hard cores of the like particles is set to zero and the 
one of the unlike particles is set to $\sigma/2$, we find that the 
protons component develops a H$_2$ first peak at a radius slightly smaller
than in the case of the additive mixture with a protons hard core 
equal to $\sigma$. Moreover the second peak almost disappears in the fermions 
case. This can be explained by the fact that the electrons cloud around each 
proton produce a ``dressed'' proton. These dressed protons like to form  
H$_n$ molecules with $n=2$, slightly smaller than the H$_2$ molecules
formed in the additive mixture, and no other $n>2$ molecules, unlike the
additive mixture.

Looking ahead to the simulation of the real electron-proton plasma in nature
we must take care of the following points:
i. Use the pair Coulomb density matrix to eliminate the artificial hard core 
necessary to treat the primitive approximation; ii. Carefully asses the 
convergence towards the continuum limit. This requires to choose $\tau$
so that the standard deviations for the free particle diffusions of the two 
species be both much smaller than the size of the simulation box,
i.e. $\sigma_i=\sqrt{\tau/\mu_i}\ll L$, $i=1,2$
\footnote{For example in the simulation of Fig. \ref{fig:gijH2} 
and \ref{fig:gijH2na} we have $L=5$ and 
$\sigma_e=3.16228,\sigma_p=0.0737982$. So this condition is not 
really satisfied.}
; iii. Choose carefully the thermodynamic state of temperature and density
to reproduce. Being it for a planet interior like the one for Jupiter 
\cite{Saumon2004} or 
for earthly laboratory conditions like the one of high pressures under 
diamond anvil cells
\cite{Hemley1988}. 
 
\appendix
%%%%%%%%%%%%%%%%%%%%%%%%%%%%%%%%%%%%%%%%%%%%%%%%%%%%%%%%%%%%%%%%%%%%%%%%%%%%%%
\section{Description of our PIMC and RPIMC algorithms}
%%%%%%%%%%%%%%%%%%%%%%%%%%%%%%%%%%%%%%%%%%%%%%%%%%%%%%%%%%%%%%%%%%%%%%%%%%%%%%
\label{app:alg}

Our PIMC algorithm is made of two kinds of moves: a single slice move 
and a multislice move \cite{Ceperley1995}. In the single slice move we 
perform a uniform displacement of a single particle coordinate at a given 
timeslice. We accept or reject the move according to the Metropolis algorithm
\cite{Metropolis,Kalos-Whitlock}. In the multislice move we create two 
Brownian bridges \cite{Kalos-Whitlock} between the initial positions 
taken from the paths of two particles $\alpha$ and $\gamma$ at the same 
timeslice and the final positions taken from the same two paths at a 
subsequent timeslice but exchaged, $\gamma$ and $\alpha$, so to connect 
particle $\alpha$ at the initial timeslice to particle $\gamma$ at the 
final timeslice with one bridge and particle $\gamma$ at the initial 
timeslice with particle $\alpha$ at the final timeslice. Again we 
accept or reject the move according to Metropolis algorithm. If the move
is accepted one creates an exchange of two particles. Since any permutation
of $N_i$ particles can be obtained by composing a finite number of like 
particles exchanges this is sufficient to simulate Bose-Einstein statistics
in an exact numerical way.
For Fermi-Dirac statistics one will face the ``fermions sign problem''
\cite{Ceperley1991}. In order to overcome this it is necessary to use
an approximate algorithm. We chose the RPIMC described in Ref. 
\cite{Ceperley1996} enforcing a free particle (approximate) restriction 
during the simulation.

The listing of our FORTRAN code is given in the supplemental material.

%%%%%%%%%%%%%%%%%%%%%%%%%%%%%%%%%%%%%%%%%%%%%%%%%%%%%%%%%%%%%%%%%%%%%%%%%%%%%%
%%%%%%%%%%%%%%%%%%%%%%%%%%%%%%%%%%%%%%%%%%%%%%%%%%%%%%%%%%%%%%%%%%%%%%%%%%%%%%
%%%%%%%%%%%%%%%%%%%%%%%%%%%%%%%%%%%%%%%%%%%%%%%%%%%%%%%%%%%%%%%%%%%%%%%%%%%%%%

\section*{Author declarations}

\subsection*{Conflicts of interest}
None declared.

\subsection*{Data availability}
The data that support the findings of this study are available from the 
corresponding author upon reasonable request.

\subsection*{Funding}
None declared.

%%%%%%%%%%%%%%%%%%%%%%%%%%%%%%%%%%%%%%%%%%%%%%%%%%%%%%%%%%%%%%%%%%%%%%%%%%%%%%
\bibliography{eipimc}
%\bibliographystyle{prsty}

%%%%%%%%%%%%%%%%%%%%%%%%%%%%%%%%%%%%%%%%%%%%%%%%%%%%%%%%%%%%%%%%%%%%%%%%%%%%%%
%%%%%%%%%%%%%%%%%%%%%%%%%%%%%%%%%%%%%%%%%%%%%%%%%%%%%%%%%%%%%%%%%%%%%%%%%%%%%%
%%%%%%%%%%%%%%%%%%%%%%%%%%%%%%%%%%%%%%%%%%%%%%%%%%%%%%%%%%%%%%%%%%%%%%%%%%%%%%
\end{document}